# A new interpretation of quantum mechanics

V. A. Golovko

Moscow State University of Mechanical Engineering (MAMI)
Bolshaya Semenovskaya 38, Moscow 107023, Russia
E-mail: fizika.mgvmi@mail.ru

Abstract

The present paper is based upon equations obtained in an earlier paper by the author devoted to a new formulation of quantum electrodynamics. The equations describe the structure of the electron as well as its motion in external fields, interaction with a measuring apparatus inclusive, in a deterministic manner without any jumps. Quantum mechanics is an approximate theory because its equations follow from the above equations upon neglecting the self-field of the electron itself. Just this leads to paradoxes, seeming contradictions and jumps in quantum mechanics. The quantum mechanical wavefunction has a dual interpretation. In some problems the square of its modulus represents a real distribution of the electronic density while in others the same square determines the probability distribution of coordinates. It is shown why, given the different interpretations of the wavefunction, it satisfies one and the same Dirac or Schrödinger equation. Description of many-electron systems is also considered in the starting approach as well as in quantum mechanics. Neutrinos are discussed in brief.

# 1. Introduction

Discussions about the interpretation of quantum mechanics (QM) and in particular about the meaning of the wave function continue up till now. The author holds to the opinion that the interpretation of the wave function can be differing in different situations. In this respect the author agrees with Barut and his co-workers [1]. At the same time it must be underlined at once that the interpretation of QM proposed in the present paper differs cardinally from the one that was elaborated by Barut and his co-workers.

Problems solved in QM can be divided into two classes. To the first class of problems can be assigned, for example, the problem of the structure of an atom (for simplicity's sake we shall imply a hydrogen atom). Once we presume that atoms exist irrespective of whether or not we observe them, at a given moment of time the electron must be somewhere in the atom. Here the probabilistic interpretation of the wave function $\psi$ is inappropriate and the quantity $|\psi|^2$ should represent a real distribution of the electronic density in the atom. This is a point of view Schrödinger adhered to initially [1]. However difficulties emerge here straightaway. First, the electronic matter creates an electromagnetic field that must act upon the matter itself giving rise to an energy of self-interaction of the electronic field. If, however, this energy is directly introduced into the Schrödinger equation, the results obtained will be in drastic contradiction with experiment [2]. Secondly, in the case of many-electron systems the wave function depends upon coordinates of each electron: $\psi = \psi(\mathbf{r}_1, \mathbf{r}_2, \mathbf{r}_3, \ldots)$. Here the interpretation of the wave function as a genuine field in space cannot hold. By and large the first class comprises problems where one deals with individual quantum systems and their evolution in time.

To the second class of problems can be assigned, for example, scattering of particles by a field of force or diffraction of particles. One implies here that the experiment is carried out with a large number of mutually independent particles. In this instance the statistical interpretation of the wave function is natural, and in all textbooks on QM one implies just this class of problems when discussing the probability interpretation of the wave function. It is worth remarking that a similar class of problems is considered in classical mechanics too where statistical methods are employed as well, and one can even point out hidden variables: they are impact parameters and initial velocities of particles. The famous Rutherford formula of 1911 for the scattering cross-section was derived by means of classical mechanics.

The question that arises in this connection is why such different problems as those of first and second classes are treated on the basis of one and the same Schrödinger equation. On the other hand, it should be pointed out that one takes fundamentally different solutions to that equation: one utilizes solutions vanishing rapidly at infinity for the first class of problems (the integral $\int_{(\infty)} |\psi|^2 \, dV$ converges) whereas for the second class of problems are utilized solutions that do not decrease or decrease slowly at infinity (the integral $\int_{(\infty)} |\psi|^2 \, dV$ diverges).

In Ref. [3] a new formulation of quantum electrodynamics (QED) was proposed in which the electronic and electromagnetic fields are ordinary *c*-numbers in contradistinction to noncommuting *q*-numbers used in the standard formulation of QED. The fact that the wave function (represented by two bispinors in that case) is a *c*-number enables one to elucidate the structure of the electron and investigate its behavior in external fields. This provides a means of solving the first-class problems. The motion of the electron ought to proceed in a continuous manner in space and time and no jumps like the wavefunction collapse at the moment of measurement can occur here, which will be considered in Sec. 2. In the subsequent sections upon passing from QED to QM we shall see why such jumps and other paradoxes make their appearance in QM. We shall also provide answers to difficulties and questions formulated above.

## 2. Motion of the electron

Let us write down the basic equations obtained in [3]. The electron is described by two bispinors $\psi_1$ and $\psi_2$ that satisfy two mutually connected Dirac equations

$$ic\hbar\gamma^\mu \frac{\partial \psi_1}{\partial x^\mu} - e\left(A_\mu + A_\mu^{\text{ext}}\right)\gamma^\mu \psi_1 - \frac{\hbar c}{\lambdabar}\psi_1 - ev_\mu\gamma^\mu \psi_2 = 0, \tag{2.1}$$

$$ic\hbar\gamma^\mu \frac{\partial \psi_2}{\partial x^\mu} - e\left(A_\mu + A_\mu^{\text{ext}}\right)\gamma^\mu \psi_2 - \frac{\hbar c}{\lambdabar}\psi_2 + ev_\mu\gamma^\mu \psi_1 = 0. \tag{2.2}$$

Here and henceforth, if the opposite is not pointed out, we imply the definitions and notation adopted in [4]; $c$, $\hbar$ and $e$ are the standard constants. Besides, $A_\mu^{\text{ext}}$ is an external four-potential and the third terms contain, instead of the electron mass *m*, a combination including the electron Compton wavelength $\lambdabar$ related to *m* by Eq. (5.2) of [3]. The four-vector $v_\mu$ that connects the equations is specified by the condition

$$\overline{\psi}_1 \gamma^\mu \psi_2 + \overline{\psi}_2 \gamma^\mu \psi_1 = 0. \tag{2.3}$$

The self electromagnetic field is determined by the Maxwell equations for the four-potential $A^\mu$ that are of the form

$$\frac{\partial F^{\mu\nu}}{\partial x^\nu} = -4\pi e\left(\overline{\psi}_1 \gamma^\mu \psi_1 - \overline{\psi}_2 \gamma^\mu \psi_2\right), \tag{2.4}$$

where $F^{\mu\nu} = \partial A^{\nu}/\partial x_{\mu} - \partial A^{\mu}/\partial x_{\nu}$ is the electromagnetic field tensor. The equations are supplemented with the relativistically invariant normalization condition

$$\int\limits_{(\infty)} (\psi_1^* \psi_1 - \psi_2^* \psi_2) dV = 1. \quad (2.5)$$

All quantities in the above equations are classical *c*-numbers.

The equations depict the behavior and movement of the electron in external electromagnetic fields which can be considered to be continuous. As long as the electronic field represented by the bispinors $\psi_1$ and $\psi_2$ is a real material substance, it can evolve in space and time only in a continuous manner without any jumps. Moreover, the equations are relativistically invariant so that all processes that occur with the electron satisfy the requirements of special relativity that does not allow for instantaneous processes.

It should be emphasized that the electron mass *m* does not figure in the above equations. According to [3] the mass is a non-relativistic notion and comes into being in the non-relativistic approximation alone. The electron mass *m* can also be introduced when the external field far exceeds the electronic self-field so that one can put $A_\mu = v_\mu = 0$ in a first approximation. Then the bispinor $\psi_1$ solely will figure in (2.1) while one ought to take $\psi_2 = 0$ for the solution to Eq. (2.2). The neglect of electronic self-field signifies that one should put $\alpha = 0$ in Eq. (5.2) of [3], which gives $m = \hbar/\lambda c$ and therefore $\hbar c/\lambda = mc^2$. Substituting this into the third summand in (2.1) yields the ordinary Dirac equation for the electron in the external field $A_\mu^{\text{ext}}$.

If this Dirac equation without the electronic self-field has a spatially bounded solution, that is to say, the integral in (2.5) at $\psi_2 = 0$ exists with this $A_\mu^{\text{ext}}$, then one may look for corrections to the solution due to the electronic self-field in terms of expansions in powers of the fine-structure constant $\alpha = e^2/\hbar c$. To this end one is to introduce the dimensionless quantities as in (4.1) of [3] and to reduce Eqs. (2.1) and (2.2) to the form of Eqs. (4.2) and (4.3) of [3]. Seeing that Eqs. (2.1) and (2.2) were derived from standard QED equations, we must obtain the same corrections as in QED. So, for example, one may find the Lamb shift of levels in a hydrogen atom. Therein lies the answer to the first difficulty mentioned in Introduction, the difficulty relevant to the energy of self-interaction of the electronic field: the energy is correctly taken into account in QED and leads to a experimentally observed effect, the Lamb shift in the present case. It must be underlined that the corrections in terms of expansions in powers of $\alpha$ can be sought only for the first-class problems (see Introduction) and only when the Dirac equation without regard for the electronic self-field has solutions localized in space.

The equations obtained in [3] were analyzed in detail only for a free electron at rest. Such equations can be solved but numerically. Upon solving the equations for the electron at rest one is able to find out the form of the electron moving at a constant speed by passing merely into a moving coordinate system with the help of the Lorentz transformation. Inasmuch as the electron at rest does not spread in space, the moving electron will not spread either [Eq. (2.5) will hold as before] as distinct from spreading wave packets in QM. Forces that keep the electron from spreading are pointed out in [3]. According to [3] the free electron at rest has the size lying in the range between the electron Compton wavelength and the Bohr radius in a hydrogen atom, and resembles a cloud. When meeting with a proton the electronic cloud envelops it, thus giving rise to a hydrogen atom. It might be observed in passing that the size of the electronic cloud increases in this event (up to the Bohr radius) in spite of the attraction to the proton.

Among four fundamental interactions – gravitational, electromagnetic, strong and weak – in the framework of standard QM it is worthwhile to speak of the electromagnetic interaction alone. By virtue of this, all processes and phenomena studied in QM can be considered on the basis of Eqs. (2.1)–(2.5) by choosing the required external potential $A_\mu^{ext}$ properly. As a famous and instructive example let us discuss the passage of an electron through two closely spaced slits cut in an impenetrable screen. It should be emphasized at once that in order to solve this problem exactly one needs to know the atomic structure of the screen and to find the relevant potential $A_\mu^{ext}$. The electronic cloud passes through both the slits simultaneously but, nevertheless, represents the single electron as before owing to the normalization condition of (2.5) that conserves its form regardless of fields acting on the electron. After passing through the slits, when the field of the screen ceases to act, the electron restores its previous form of a free electron because this form is unique. If one knows the initial location and velocity of the electron and the atomic structure of the screen as well, one is able to calculate absolutely exactly the direction in which the electron will move after passing through the slits and its location at an arbitrary moment of time. It will be noted that the electron size mentioned above is only a conventional size of the electronic cloud and the electron wave function extends far beyond the electron size so that the electron is sensitive to objects rather remote from its centre. The example considered demonstrates also that the electronic cloud can have diverse forms, and even disruptions in it can form under the influence of external fields but the electron restores its form in the wake of cessation of external actions. We shall yet revert to the example later on.

In an analogous way one may consider interaction of the electron with a measuring apparatus. The measuring apparatus is also described by a potential $A_\mu^{ext}$ that can depend upon the time too; besides, one must know the manner in which the electron acts on the apparatus. Upon solving Eqs. (2.1)–(2.5) we shall find the exact location and velocity of the electron after the measurement and the state of the measuring apparatus. In much the same way one can analyze the Einstein-Podolsky-Rosen paradox although two particles are involved there whereas the case of several particles will be considered in Sec. 4. But the essence remains the same: we are capable of calculating exactly the state of the particles after the collision and what happens when one of the particles is acted upon by a measuring instrument. Of course to carry out exact calculations on the basis of Eqs. (2.1)–(2.5) for the examples discussed in this and the preceding paragraphs is practically impossible. At the same time QM enables one to perform some calculations in those cases too, from which it becomes clear that the calculations will be approximate and connected with loss of some information.

### 3. Passage to quantum mechanics

QM differs from QED in that the former does not take the electronic self-field into account while considering the motion of an electron in external fields. In the previous section we discussed the case where the electronic self-field could be ignored as compared to $A_\mu^{\text{ext}}$. The reasoning was relevant to one-electron problems of first class mentioned in Introduction. In the present section we turn to the second class of problems.

One implies in the second-class problems that an experiment is carried out with a large number of mutually independent particles. Let there be a system composed of an electron that moves in a given external field, and we do not imply any simplification for the moment. We introduce now an ensemble by analogy with statistical mechanics. The ensemble will comprise a very large number $\mathcal{N}$ of systems each of which is a replica of the starting system distinct from one another in the initial position and velocity of the electron. Let $\psi^{(i)}(x)$, $i = 1,\ldots, \mathcal{N}$, be the wave function of each electron in the ensemble where $x$ denotes for short the set of coordinates and time ($x^\mu$, $\mu = 0,1,2,3$). The function $\psi^{(i)}(x)$ may be the bispinor $\psi_1$ (for the sake of simplicity we put $\psi_2 = 0$ at once because we shall later on proceed to the limit $A_\mu \to v_\mu \to 0$ when $\psi_2 \to 0$ according to the previous section) or it may be a non-relativistic wave function. Furthermore, we introduce the function

$$\Psi(x) = \frac{1}{\sqrt{\mathcal{N}}} \sum_{i=1}^{\mathcal{N}} \psi^{(i)}(x). \qquad (3.1)$$

If $\psi^{(i)}(x)$ is the bispinor, $\Psi(x)$ will be a bispinor as well.

The ensemble will be prepared so that the functions $\psi^{(i)}(x)$ do not overlap. This can be done because the functions $\psi^{(i)}(x)$ drop exponentially from the centre [3]. In this case

$$\int \psi^{(i)*}(x)\psi^{(k)}(x)dV = \delta_{ik}, \qquad (3.2)$$

where $\delta_{ik}$ is the Kronecker symbol and (2.5) with $\psi_2 = 0$ is taken into account if $i = k$. The integration in (3.2) is carried out over any volume exceeding the size of the particles provided the particle under number $i = k$ is completely located in the volume, otherwise the integral equals zero (we imply such volumes when it can be neatly stated whether or not the particle is in the volume). In order to take all situations into consideration one may assume that the initial functions $\psi^{(i)}(x)$ cover the space in question more or less uniformly. In the course of the motion the functions can overlap. Such events, however, will be comparatively infrequent and they should not make a noticeable contribution to quantities calculated if $\mathcal{N}$ is large and the initial positions of the particles in the ensemble are widely spaced.

Let there be $\Delta\mathcal{N}$ particles of the ensemble in a volume $\Delta V$. Then with use made of (3.2) we obtain

$$\int_{\Delta V} \Psi^*(x)\Psi(x)dV = \frac{\Delta\mathcal{N}}{\mathcal{N}}. \tag{3.3}$$

This result demonstrates that $\Psi(x)$ possesses properties of a wave function if the last is interpreted statistically. If the integration in (3.3) is carried out over all space, then

$$\int_{(\infty)} \Psi^*(x)\Psi(x)dV = 1, \tag{3.4}$$

that is to say, one obtains the standard normalization of the wave function. The particles of the ensembles can fill up all space. Then $\mathcal{N} \to \infty$. In this case the integral in (3.4) will diverge. Such a case is also encountered in QM.

We continue with investigating the function $\Psi(x)$ in the spirit of QM. Let there be a physical quantity the operator $\widehat{F}$ corresponds to (as in [3] we mark operators with an arc at the top). With use made of (3.2) we calculate the integral

$$\int_{(\infty)} \Psi^*(x)\widehat{F}\Psi(x)dV = \frac{1}{\mathcal{N}}\sum_{i=1}^{\mathcal{N}} F^{(i)}, \tag{3.5}$$

wherein

$$F^{(i)} = \int_{(\infty)} \psi^{(i)*}(x)\widehat{F}\psi^{(i)}(x)dV. \tag{3.6}$$

Seeing that the expression on the right side of (3.5) makes sense of an average value, the average value is computed with the help of the integral on the left side of (3.5) as is usual in QM.

Upon solving the equation $\widehat{F}\psi_n(x) = F_n\psi_n(x)$ we will find the eigenfunctions $\psi_n(x)$ presuming that they form a complete orthonormal set of functions [the functions $\psi_n(x)$ may be of the form $\psi_n(x) = \psi_n(\mathbf{r})\exp(-i\varepsilon_n t/\hbar)$]. Then $\psi^{(i)}(x)$ can be expanded in terms of $\psi_n(x)$:

$$\psi^{(i)}(x) = \sum_n a_n^{(i)} \psi_n(x). \tag{3.7}$$

Substituting this into (3.6) and (3.5) yields

$$\int_{(\infty)} \Psi^*(x)\widehat{F}\Psi(x)dV = \frac{1}{\mathcal{N}} \sum_{i=1}^{\mathcal{N}} \sum_n F_n \mid a_n^{(i)} \mid^2 . \tag{3.8}$$

We define average values $\bar{a}_n$ according to

$$\mid \bar{a}_n \mid^2 = \frac{1}{\mathcal{N}} \sum_{i=1}^{\mathcal{N}} \mid a_n^{(i)} \mid^2 . \tag{3.9}$$

Introducing this into (3.8) we shall find that

$$\int_{(\infty)} \Psi^*(x)\widehat{F}\Psi(x)dV = \sum_n F_n \mid \bar{a}_n \mid^2. \tag{3.10}$$

The function $\Psi(x)$ can be directly expanded by analogy with (3.7)

$$\Psi(x) = \sum_n a_n \psi_n(x). \tag{3.11}$$

Calculating the integral on the left of (3.10) with this $\Psi(x)$ we shall obtain agreement with (3.10) on condition that $|a_n| = |\bar{a}_n|$. Consequently, instead of (3.1) one can employ a usual for QM expression of the type (3.11).

We discuss now how the energy and momentum can be calculated in the present situation. The density of energy $T^{00}$ is given by Eq. (3.19) of [3]. Upon neglecting the energy of the electronic self-field, that is, upon putting $E = H = \varphi = 0$ and $\psi_2 = 0$ as well and replacing $\psi_1$ by $\psi^{(i)}(x)$ we obtain

$$T_{(i)}^{00} = \frac{i\hbar}{2}\left(\psi^{(i)*}\frac{\partial \psi^{(i)}}{\partial t} - \frac{\partial \psi^{(i)*}}{\partial t}\psi^{(i)}\right). \tag{3.12}$$

Therefore the energy of one particle in the ensemble is

$$\mathcal{E}^{(i)} = \int_{(\infty)} T_{(i)}^{00} dV = \frac{i\hbar}{2}\int_{(\infty)}\left(\psi^{(i)*}\frac{\partial \psi^{(i)}}{\partial t} - \frac{\partial \psi^{(i)*}}{\partial t}\psi^{(i)}\right)dV . \tag{3.13}$$

The energy of one particle averaged over the ensemble will be

$$\mathcal{E} = \frac{1}{\mathcal{N}}\sum_{i=1}^{\mathcal{N}} \mathcal{E}^{(i)} = \frac{i\hbar}{2}\frac{1}{\mathcal{N}}\sum_{i=1}^{\mathcal{N}}\int_{(\infty)}\left(\psi^{(i)*}\frac{\partial \psi^{(i)}}{\partial t} - \frac{\partial \psi^{(i)*}}{\partial t}\psi^{(i)}\right)dV . \tag{3.14}$$

With use made of (3.2) and of (3.1) afterwards this can be recast as

$$\mathcal{E} = \frac{i\hbar}{2}\frac{1}{\mathcal{N}}\sum_{i,k=1}^{\mathcal{N}}\int_{(\infty)}\left(\psi^{(i)*}\frac{\partial\psi^{(k)}}{\partial t} - \frac{\partial\psi^{(i)*}}{\partial t}\psi^{(k)}\right)dV = \frac{i\hbar}{2}\int_{(\infty)}\left(\Psi^*\frac{\partial\Psi}{\partial t} - \frac{\partial\Psi^*}{\partial t}\Psi\right)dV. \qquad (3.15)$$

Thus we have expressed the mean energy of the particle in terms of $\Psi(x)$. As a check we take a stationary state where the wave function is to be of the form $\Psi(x) = \psi(\mathbf{r})\exp(-i\varepsilon t/\hbar)$. If this is substituted into the last formula of (3.15), we obtain $\mathcal{E} = \varepsilon$, which amounts to saying that the quantity $\varepsilon$ in the expression $\psi(\mathbf{r})\exp(-i\varepsilon t/\hbar)$ is the energy of the particle, as it should. It may be remarked that the formula of (3.15) does not include the relativistic rest energy $mc^2$, although the exact formula for the energy of the electron (4.27) of [3] does include it. The point is that upon putting $\varphi = 0$ we have discarded the summand $-1/\alpha$ in (4.26) of [3] as well whereas just this summand corresponds to the energy $mc^2$.

We are coming next to the momentum. The formula for the momentum was not written down in Ref. [3]. It can be obtained with use made of the energy-momentum tensor (3.18) of [3] and with account taken of the fact that the components of the momentum density are $T^{10}/c$, $T^{20}/c$ and $T^{30}/c$ [5]. As a result, we find for the momentum of the electron $\mathbf{P}$ in the three-dimensional notation of [3] that

$$\mathbf{P} = \frac{1}{c}\int_{(\infty)}\left\{\frac{1}{4\pi}[\mathbf{E}\mathbf{H}] - e\mathbf{A}\left(\psi_1^*\psi_1 - \psi_2^*\psi_2\right)\right\}dV - i\hbar\int_{(\infty)}\left(\psi_1^*\nabla\psi_1 - \psi_2^*\nabla\psi_2\right)dV. \qquad (3.16)$$

It is worthy of remark that the formula was derived under the assumption that the motion of the electron is finite, that is, upon discarding surface integrals at infinity. If one ignores the electronic self-field as in (3.12) and replaces $\psi_1$ by $\psi^{(i)}(x)$, one obtains from (3.16)

$$\mathbf{P}^{(i)} = -i\hbar\int_{(\infty)}\psi^{(i)*}\nabla\psi^{(i)}dV. \qquad (3.17)$$

We proceed further as in passing from (3.13) to (3.15). As a result

$$\mathbf{P} = -i\hbar\int_{(\infty)}\Psi^*\nabla\Psi dV. \qquad (3.18)$$

Comparing this with (3.5) we see that in the case under consideration the momentum operator is $-i\hbar\nabla$ as is usual in QM.

Thus we see that the function $\Psi(x)$ satisfies basic relations concerning the wave function in QM. It remains to find the equation for the function. As stated at the outset of the present section QM does not take the electronic self-field into account. For this reason in order to find the equation for the functions $\psi^{(i)}(x)$ that enter into the definition of (3.1) it suffices to put $A_\mu = v_\mu = 0$ in (2.1). The resulting equation proves linear so that the superposition of solutions that is the function $\Psi(x)$ according to (3.1) will satisfy the same equation. Hence

$$ic\hbar\gamma^\mu \frac{\partial \Psi}{\partial x^\mu} - eA_\mu^{ext} \gamma^\mu \Psi - mc^2 \Psi = 0 \,. \tag{3.19}$$

Here we have introduced the electron mass according to $m = \hbar/\lambdabar c$ instead of $\lambdabar$, which can be done when $A_\mu = v_\mu = 0$ (see the previous section). Equation (3.19) is the standard Dirac equation. Upon passing to the non-relativistic limit in (3.19) one obtains the Schrödinger equation.

The following remark should be made at this point. According to the starting definition of (3.1) the function $\Psi(x)$ is not at all smooth but represents a set of peaks situated in the maxima of the functions $\psi^{(i)}(x)$, the peaks being generally distributed at random. It should be stressed that all formulae from (3.3) to (3.18) remain valid in this instance too. Equation (3.19), however, has not such solutions: its solutions are smooth given the sufficient smoothness of the potential $A_\mu^{ext}$. For this reason, when passing to Eq. (3.19) we factually carry out smoothing over the irregularities of $\Psi(x)$. In reality the smoothing occurs automatically because of the spreading of wave packets characteristic of QM. Only taking account of the self-field holds the wave packet relevant to the electron from spreading.

In summary, we find the answer to the question formulated in Introduction as to why the wave function satisfies one and the same equation given its different interpretation for the problems of first and second classes. Of course, for the first-class problems one could also use the statistical interpretation necessary for the second-class problems by considering, for instance, an ensemble of a large number of atoms. But this obscures the problem, and nothing else.

In essence, the second class embraces the problems where the Schrödinger equation (for brevity, we shall speak of that equation alone) has no localized solutions. Such solutions would exist on account of the electronic self-field but QM ignores this field. As long as the electron cannot fill up the whole of space, it is clear that the probabilistic interpretation of the wave function alone can be implied in this situation. If, nevertheless, one tries to speculate about the trajectory of an individual electron or about interaction of the electron with a measuring apparatus, then there appear paradoxes and misunderstandings. The added complication is that QM does in no way take the sizes of the electron into account. Factually, the electron is reckoned as a point-like particle. When one states that the wave function determines the probability of finding a particle in a volume element, the particle is thought of as a point because one can exactly say solely about the point whether or not it is in the volume element.

We revert to discussing the double-slit experiment. In Sec. 2 we have established that the electronic cloud deforms and passes through both the slits simultaneously. Insofar as the electron is regarded as a point in QM, the question arises there as to through which of the slits the electron passes seeing that the point cannot pass through both the slits simultaneously. So long as the electron is not deformed by an external action, it can, in certain respects, be considered to be the point. The electron deforms owing to interaction with the substance of the screen. Just this process is not taken into consideration in QM. One may say that the screen with the surrounding field is factually represented in QM like a structureless surface. One can here draw a parallel with geometrical optics. According to geometrical optics, when light is reflected, its velocity instantaneously changes the direction. To the instantaneous variation of the velocity corresponds, however, an infinite acceleration. Therefore, from the point of view of geometrical optics the reflection of light must provoke an infinite force acting on the reflecting surface. This paradox of geometrical optics is never discussed, however, because no paradoxes occur when use is made of the undulatory theory of light. The situation in QM is, of course, much more complicated as compared with geometrical optics but all potential barriers and wells are represented in QM as structureless objects. On the one hand, this essentially simplifies calculations (cf. the discussion at the end of Sec.2) while on the other hand the inevitable in this case probabilistic approach leads to loss of some information and to the appearance of seeming paradoxes. To the advantages of QM may be added the fact that it enables one to consider the motion of any particles that can be thought of as structureless, not only the motion of the electron.

From the structure of the electron studied in [3] it is clearly seen that the electron is not a wave but it is a particle pure and simple though with a complex constitution. Hence there is no particle-wave duality of the electron. At the same time, upon deforming the electronic cloud is able to bend around an obstacle like the wave does. As a consequence, when the experiment is performed with a large number of electrons (the second-class problems), one observes phenomena akin to the diffraction of waves. Separate electrons pass through the slits in various ways depending on their initial positions and velocities. Upon averaging there appears a diffraction pattern.

Since the integrand in (2.5) has a meaning of the density of the electronic matter, the centre of the electron can be defined as

$$\mathbf{R} = \int_{(\infty)} \mathbf{r}(\psi_1^* \psi_1 - \psi_2^* \psi_2) dV \ . \tag{3.20}$$

The momentum of the electron **P** is determined uniquely by Eq. (3.16). We have no uncertainty relation here. As to QM where the electron is reckoned a point the location of the point can be anywhere in the domain where the electronic density $\psi_1^* \psi_1 - \psi_2^* \psi_2$ is markedly different from zero. Inasmuch as the electronic matter rotates as indicated by a nonzero spin and moreover the matter can deform, the velocities of its diverse parts are different. Therefrom come about the uncertainty relations which make sense only if the electron is thought of as a point.

## 4. Systems with several electrons

Up till now we discussed properties and behavior of an individual electron. Ref. [3] is also devoted to the individual electron alone. Let us elucidate how one can treat systems containing several electrons along the lines of [3]. Ref. [3] still includes a discussion of the simultaneous consideration of two particles, namely, of an electron and a positron. Both the particles are described by a unified $\psi$-field satisfying Eqs. (2.1)–(2.4) but in the region where the electron is initially located we have the condition of (2.5) while in the one of the initial location of the positron we have the same condition with $-1$ on the right now. This example indicates that several electrons are describable by a single electronic field satisfying Eqs. (2.1)–(2.4) just as one electron is. The sole distinction will be in that, if there are $N$ electrons, instead of (2.5) one ought to take

$$\int_{(\infty)} (\psi_1^* \psi_1 - \psi_2^* \psi_2) dV = N . \tag{4.1}$$

All of this can be illustrated by the example a many-electron atom. The electronic envelope of the atom is described by a common unified electronic field where there are no individual electrons. If the atom is ionized, a part of the electron cannot escape from the atom, but one electron as a single whole leaves the atom inasmuch as the condition of (2.5) must hold for a free electron. If the atom is ionized completely, $N$ electrons escape from the atom, and we state that the atom consisted of $N$ electrons. Figuratively speaking, the individual electrons are created when quitting the atom. Likewise, the electron can be absorbed by the atom only completely. Of course, inside the electronic envelope there can be clots reminiscent of separate electrons but this can be established solely upon solving Eqs. (2.1)–(2.4) with the condition of (4.1), which can be done only numerically. It is worthy of remark that in the standard theory of a many-electron atom it is practically impossible to trace the path of each electron. When one speaks of a state of an individual electron in the atom, one implies the state of the electron in some effective centrally symmetric field created by the nucleus together with all the other electrons, the electron moving independently of the others (see, e.g., [6]). Such an approximation proves to be astonishingly good.

In this connection the following question may be raised. It was shown in [3] that the electron spin is equal to $M_z = \hbar I/2$ where $I$ is the integral on the left side of (4.1). For this reason it would seem that the spin of the above electronic envelope of the atom should always equal $M_z = \hbar N/2$ in view of (4.1). At the same time the spin of the electronic envelope of real atoms can be different and, when $N$ is large, it can be considerably less than $\hbar N/2$. We shall prove, however, that the total spin of several electrons is not uniquely related to the normalization condition of the type (4.1). It is demonstrable that in the case of axially symmetric solutions the bispinors $\psi_1$ and $\psi_2$ can have a form more general than in (4.4) of [3], namely,

$$\psi_1 = \begin{pmatrix} f_1^{(1)}(\rho,z)e^{i\nu\dot\varphi} \\ f_2^{(1)}(\rho,z)e^{i(\nu+1)\dot\varphi} \\ if_3^{(1)}(\rho,z)e^{i\nu\dot\varphi} \\ if_4^{(1)}(\rho,z)e^{i(\nu+1)\dot\varphi} \end{pmatrix}, \quad \psi_2 = \begin{pmatrix} f_1^{(2)}(\rho,z)e^{i\nu\dot\varphi} \\ f_2^{(2)}(\rho,z)e^{i(\nu+1)\dot\varphi} \\ if_3^{(2)}(\rho,z)e^{i\nu\dot\varphi} \\ if_4^{(2)}(\rho,z)e^{i(\nu+1)\dot\varphi} \end{pmatrix}, \quad (4.2)$$

where $\dot\varphi$ is the angle in cylindrical coordinates. The formulae of (4.4) of [3] follow therefrom if $\nu = 0$. According to Dirac's argumentation [7], § 36, the absolute value of the number $\nu$ must be an integer (zero inclusive) or a half-integer. The argumentation, however, is inapplicable to our case because it is relevant to QM which does not describe the electronic self-field. As before, the physical quantities will not depend on $\dot\varphi$, the vector **A** will have only one component $A_\varphi$ and div **A** = 0. The spin of the electronic formation can be calculated by Eq. (3.21) of [3]. As in the case of a single electron the first integral in that equation vanishes for stationary solutions provided that the electric and magnetic fields decrease in the ordinary way at infinity. Substituting (4.2) into the second integral and taking (4.1) into account yields

$$M_z = \left(\frac{1}{2} + \nu\right)\hbar N. \quad (4.3)$$

We see from this that the spin of the electronic formation with a given $N$ in the ground state can be different depending on the number $\nu$. The number has to be found when solving the equations.

We turn now to the question formulated in Introduction as to why a system containing $N$ electrons is described in QM by a wave function $\psi = \psi(\mathbf{r}_1, \mathbf{r}_2, \ldots, \mathbf{r}_N)$ that depends upon several arguments. First of all it must be observed that starting from the standard Dirac equation it is impossible to strictly formulate an equation analogous to the Schrödinger equation for $N$ bodies and the relativistic problem of $N$ bodies is treated with use made of one or other of approximations [8]. This being so, we proceed from the fact that in the non-relativistic limit Eq. (2.1) yields the Schrödinger equation. In this limit the electronic self-field disappears off the equation and $A_\mu^{\text{ext}}$ alone remains [3]. When, however, we have several electrons (we use the habitual terminology), the self-energy of the unified electronic field must be taken into account, if only approximately.

In line with the foregoing it is natural to suppose that, in the unified electronic field, to separate electrons correspond clots which should be most pronounced when they are widely separated. Besides, the clots must repel. As a result, the wave function should be of the form

$$\psi(\mathbf{r},t) = \sum_{i=1}^{N} D_i[\mathbf{r} - \mathbf{r}_i(t), t]. \tag{4.4}$$

In this formula the functions $D_i(\mathbf{r})$ have a maximum at $\mathbf{r} = 0$ and fall off steeply with increasing $|\mathbf{r}|$. In QM where the particles are thought of as points (see the preceding section) one may put $|D_i(\mathbf{r})|^2 = \delta(\mathbf{r})$. We shall take the interaction energy of the clots into approximate consideration by interpreting them as points. With this aim in view, the Schrödinger equation that results in the non-relativistic limit will be written as

$$i\hbar \frac{\partial \psi}{\partial t} = -\frac{\hbar^2}{2m} \nabla^2 \psi + e[\varphi(\mathbf{r},t) + \varphi^{\text{ext}}(\mathbf{r})] \psi(\mathbf{r},t). \tag{4.5}$$

Here $\varphi(\mathbf{r},t)$ is the scalar potential of the electronic self-field and for the sake of simplicity we assume that the external scalar potential $\varphi^{\text{ext}}(\mathbf{r})$ is time independent. For simplicity's sake as well we assume that the external vector potential is absent and the magnetic self-field can be neglected. The Schrödinger equation does not explicitly contain the spin of the particle. Investigation of the motion of the spin must be performed on the basis of the exact Eqs. (2.1)–(2.4) and (4.1), which is a very complicated and separate problem. As long as the presence of the spin does not affect the behavior of the particle when the magnetic field is absent, the spin will not be taken into consideration.

We shall suppose that each of the functions $D_i$ of (4.4) satisfies an equation of the type (4.5) with its own potential:

$$i\hbar \frac{\partial D_i}{\partial t} = -\frac{\hbar^2}{2m} \nabla^2 D_i + e[\varphi_i(\mathbf{r},t) + \varphi_i^{\text{ext}}(\mathbf{r}) + C_i] D_i. \tag{4.6}$$

We have added here an unknown constant $C_i$ seeing that the potential is defined up to an arbitrary constant. Inasmuch as the function $D_i$ is delta-like, we can write approximately that

$$[\varphi_i(\mathbf{r},t) + \varphi_i^{\text{ext}}(\mathbf{r}) + C_i] D_i = [\varphi_i(\mathbf{r}_i) + \varphi^{\text{ext}}(\mathbf{r}_i) + C_i] D_i[\mathbf{r} - \mathbf{r}_i(t), t]. \tag{4.7}$$

On account of the essence of the self-field the potential $\varphi_i(\mathbf{r}_i)$ is equal to a potential created by all the other clots at the point where the clot under consideration is located:

$$\varphi_i(\mathbf{r}_i) = \sum_{\substack{k=1 \\ k \neq i}}^{N} \frac{e}{|\mathbf{r}_i - \mathbf{r}_k|}. \tag{4.8}$$

In Eqs. (4.6) and (4.7) all coordinates $\mathbf{r}_1, \mathbf{r}_2, \ldots, \mathbf{r}_N$ except for $\mathbf{r}_i$ are parameters. The constant $C_i$ can depend on these parameters. We take it in the form

$$C_i = \sum_{\substack{k=1 \\ k \neq i}}^{N} \varphi^{\text{ext}}(\mathbf{r}_k) + \sum_{\substack{k,l=1 \\ k \neq i \\ l \neq i \\ l > k}}^{N} \frac{e}{|\mathbf{r}_k - \mathbf{r}_l|}. \tag{4.9}$$

The last sum is the, divided by $e$, interaction energy of all the clots except for the clot at $\mathbf{r} = \mathbf{r}_i$. Substituting (4.8) and (4.9) into the second square brackets of (4.7) results in

$$\left[\varphi_i(\mathbf{r}_i) + \varphi^{\text{ext}}(\mathbf{r}_i) + C_i\right] = \sum_{k=1}^{N} \varphi^{\text{ext}}(\mathbf{r}_k) + \sum_{\substack{k,l=1 \\ l>k}}^{N} \frac{e}{|\mathbf{r}_k - \mathbf{r}_l|}. \tag{4.10}$$

It is important to note that the quantity on the right side of (4.10) is independent of $i$, that is, it is the same for all $i$. Given the structure of the functions $D_i$ the spatial derivatives may be written as

$$\nabla D_i = -\frac{\partial}{\partial \mathbf{r}_i} D_i[\mathbf{r} - \mathbf{r}_i(t), t] = -\nabla_i \psi, \quad \nabla^2 D_i = \frac{\partial}{\partial \mathbf{r}_i^2} D_i[\mathbf{r} - \mathbf{r}_i(t), t] = \nabla_i^2 \psi, \tag{4.11}$$

where $\nabla_i = \partial/\partial \mathbf{r}_i$.

If all of this is introduced into (4.6) and the summation over $i$ is carried out in view of (4.4), the parameter $\mathbf{r}$ disappears off the equations and we, in fact, obtain an equation for the function $\psi(\mathbf{r}_1, \mathbf{r}_2, \ldots, \mathbf{r}_N)$:

$$i\hbar \frac{\partial \psi}{\partial t} = -\frac{\hbar^2}{2m} \sum_{i=1}^{N} \nabla_i^2 \psi + e\left[\sum_{i=1}^{N} \varphi^{\text{ext}}(\mathbf{r}_i) + \sum_{\substack{i,k=1 \\ k>i}}^{N} \frac{e}{|\mathbf{r}_i - \mathbf{r}_k|}\right] \psi(\mathbf{r}_1, \mathbf{r}_2, \ldots, \mathbf{r}_N). \tag{4.12}$$

This is the standard Schrödinger equation for $N$ particles that reside in an external electric field and interact with one another via electric forces. In addition one may take into account the fact that two particles cannot be located at one and the same point simultaneously. To this end the function $\psi(\mathbf{r}_1, \mathbf{r}_2, \ldots, \mathbf{r}_N)$ ought to be antisymmetrized so that it will vanish automatically when $\mathbf{r}_i = \mathbf{r}_k$.

The presented derivation of Eq. (4.12) is, of course, approximate. However, in real situations it is impossible to solve exactly an equation of the type (4.12) even at $N = 2$ let alone the cases where $N > 2$. Because of this, one utilizes one or other of approximations (see, by way of example, the aforesaid concerning the many-electron atom). As a result, the initial approximateness of Eq. (4.12) does not manifest itself for all practical purposes. It will be recalled that the Schrödinger equation even for a single electron is approximate as long as it does not take the electronic self-field into account.

In summary, we have answered all questions formulated in Introduction, in particular, the question as to why such different problems as those of first and second classes are treated with the help of one and the same Schrödinger equation. It may be added that the interpretation of the electronic field as a real field in space is admissible in first-class problems alone and only if a single electron is concerned. If we are dealing with several electrons, the wave function $\psi(\mathbf{r}_1, \mathbf{r}_2, \ldots, \mathbf{r}_N)$ used in QM admits exclusively a statistical interpretation even for the first-class problems.

# 5. Concluding remarks

The present paper shows that the quantum mechanical wave function has a dual interpretation. In some problems the square of its modulus represents a real distribution of the electronic density in space; in others the same square determines the probability distribution of coordinates. We have elucidated in the paper why, given the different interpretations of the wave function, it satisfies one and the same Dirac or Schrödinger equation; answers to questions concerning the first interpretation were provided as well.

The motion of the electron is described in a fully deterministic manner on the basis of the QED equations obtained in [3]. In this case there are no jumps like the wavefunction collapse and the interaction with a measuring apparatus occurs in a predictable way as in classical physics. In the double-slit experiment a part of the electronic cloud passes through one of the slits while the other part through the second. Nevertheless, the electron remains to be a single entity and restores its form after passing through the slits. The behavior of the electron may be compared with the one of solitons that are met with in various branches of physics. If a soliton encounters a perturbation, the soliton too emerges from the collision unchanged. The solitons are possible only in media that are described by nonlinear differential equations while the equations obtained in [3] are nonlinear.

QM is an approximate theory because, as compared with QED, QM does not take into account the electronic self-field and the self-interaction of this field. By the way, that is why the quantum mechanical equations prove linear. It is just the neglect of that field that leads to paradoxes and seeming contradictions in QM. If the Schrödinger equation has a spatially localized solution as, for example, in the case of the electron in a hydrogen atom, the solution describes (approximately) the structure of the electron while the QED equations that take the electronic self-field into consideration yield only small corrections, in particular, the anomalous electron magnetic moment. If, however, the external field is such that the Schrödinger equation does not admit localized solutions, the probabilistic approach alone is possible then and one cannot trace the path of an individual electron in an experiment with that approach on hand. Attempts to find out from this standpoint, for instance, through which of the slits the electron passes in the double-slit experiment can lead only to misunderstandings. For the same reason there appear seeming jumps in the electron's behavior (e.g., an instantaneous wave-packet collapse at the moment of a measurement).

The electron possesses no particle-wave duality but represents a particle though with complex properties. Upon deforming the electronic cloud is able to bend around obstacles like a wave does. In consequence, when experiments are performed with a large number of electrons, one observes phenomena akin to the diffraction of waves. The uncertainty relations result solely from an approximate character of QM when the wave function is interpreted probabilistically although to an extent they reflect the fact that different parts of the electronic cloud have diverse coordinates and momenta whereas the electron in QM is thought of as a point.

When one speaks of a system with several electrons, say, of an atom, in actual truth there is a unified electronic cloud in which there are no individual electrons. The individual electrons come into existence only when a part of the electronic matter escapes from the system, solely a definite part of the matter being able to escape from it because the condition of (2.5) must hold for the departing matter. Consequently, one may state that an individual electron has left the system. Analogously, by virtue of (2.5) again the electron can penetrate into the system only completely.

Although QM does not deal usually with neutrinos, it is worthwhile to say a word about them in addition to Ref. [3]. The neutrinos correspond to solutions to Eqs. (2.1)–(2.4) relevant to objects moving at the speed of light. In the simplest case the neutrino is described by Eqs. (5.3) and (5.4) of [3] to which must be added the Maxwell equations (2.4) with the replacement $\partial/\partial t = -c\partial/\partial z$ and the condition of (2.3). A preliminary analysis of the resulting equations shows that the structure of the neutrino can be various in contradistinction to an individual electron at rest. Even the neutrino spin is not strictly definite. When a definite electron spin was obtained in Ref. [3] with the help of Eq. (3.21) of [3], it was assumed that $\partial/\partial t = 0$ and that quantities characterizing the electron were independent of the angle $\dot{\varphi}$. The neutrino created in the course of a process receives a definite spin due to the law of conservation of angular momentum. If, for example, an electron is created during the same process, the neutrino acquires a definite structure too and can be called the electron neutrino. Figuratively speaking, the neutrino is "building waste» remaining after constructing the electron with its strictly definite structure. Such a neutrino can, at a later time, come into interaction in a resonant way only with electrons. Likewise, in the course of processes involving a muon there appears a "building waste» with its own structure which can be called the muon neutrino and which is able to interact with muons alone. As was mentioned in [3], the neutrino can evolve in flight. Modifications in the neutrino structure can happen under the influence of other particles as well. The neutrino structure may change in such a manner that the neutrino will no longer be capable of interacting in the resonant way with the leptons, much less with other particles. Such neutrinos will manifest themselves in gravitational interactions alone and they are usually called the sterile neutrinos. There are experiments that count in favor of their existence whereas in other experiments aimed at their search they were not found [9, 10]. It is quite possible that just these neutrinos constitute the dark matter or its part in outer space. If this is so, the dark matter is the "industrial debris" remaining after processes with elementary particles, the debris that makes itself evident solely via gravitation.

As long as the neutrino moves at the speed of light while the mass is a non-relativistic notion [3], one cannot speak of a neutrino mass. At the same time the terms which would contain a mass *m* in the standard form of Dirac's equation remain in Eqs. (5.3) and (5.4) of [3] describing the neutrino: these are the terms with the coefficient β. Such terms may lead to effects seemingly indicative of the fact that the neutrino has a mass. A seeming neutrino mass may ensue for another reason as well. When the neutrino propagates in matter, its effective velocity $v_{\text{eff}}$ can decrease due to interaction with the particles of matter, much as the speed of light lessens in matter. The fact that $v_{\text{eff}} < c$ may be interpreted as the existence of some mass of the neutrino. All questions concerning the neutrino can be answered by solving, if only approximately, the aforementioned equations describing the neutrino upon adding there an external potential $A_\mu^{\text{ext}}$ that would characterize the action of particles forming the matter upon the neutrino.

## References

[1] A. O. Barut, *Found. Phys. Lett.* **1**, 47 (1988).

[2] L. P. Pitaevskii, *Usp. Fiz. Nauk* **163** (8), 119 (1993) [*Phys. Usp.* **36**, 760 (1993)].

[3] V. A. Golovko, arXiv: 1607.03023 [physics.gen-ph].

[4] V. B. Berestetskii, E. M. Lifshitz and L. P. Pitaevskii, Quantum *Electrodynamics* (Pergamon, Oxford 1980).

[5] L. D. Landau and E. M. Lifshitz, *The Classical Theory of Fields* (Butterworth-Heinemann, Oxford 2000).

[6] L. D. Landau and E. M. Lifshitz, *Quantum Mechanics* (Pergamon, Oxford 2000).

[7] P. A. M. Dirac, *The Principles of Quantum Mechanics* (University Press, Oxford 1967).

[8] A. O. Barut, *J. Math. Phys.* **32**, 1091 (1991).

[9] K.A. Olive *et al*. (Particle Data Group), *Chin. Phys*. C **38**, 090001 (2014), p. 235.

[10] M. G. Aartsen *et al.* (IceCube Collaboration), *Phys. Rev. Lett.* **117**, 071801 (2016).